# Computing the intrinsic grain boundary mobility tensor

Xinyuan Song[†], Liang Yang[§,*], Chuang Deng[†,*]

§ School of Aeronautical Manufacturing Engineering, Nanchang Hangkong University, Nanchang 330063, China

† Department of Mechanical Engineering, University of Manitoba, Winnipeg, MB R3T 2N2, Canada

* Corresponding authors: Chuang.Deng@umanitoba.ca (C. Deng); l.yang@nchu.edu.cn (L. Yang)

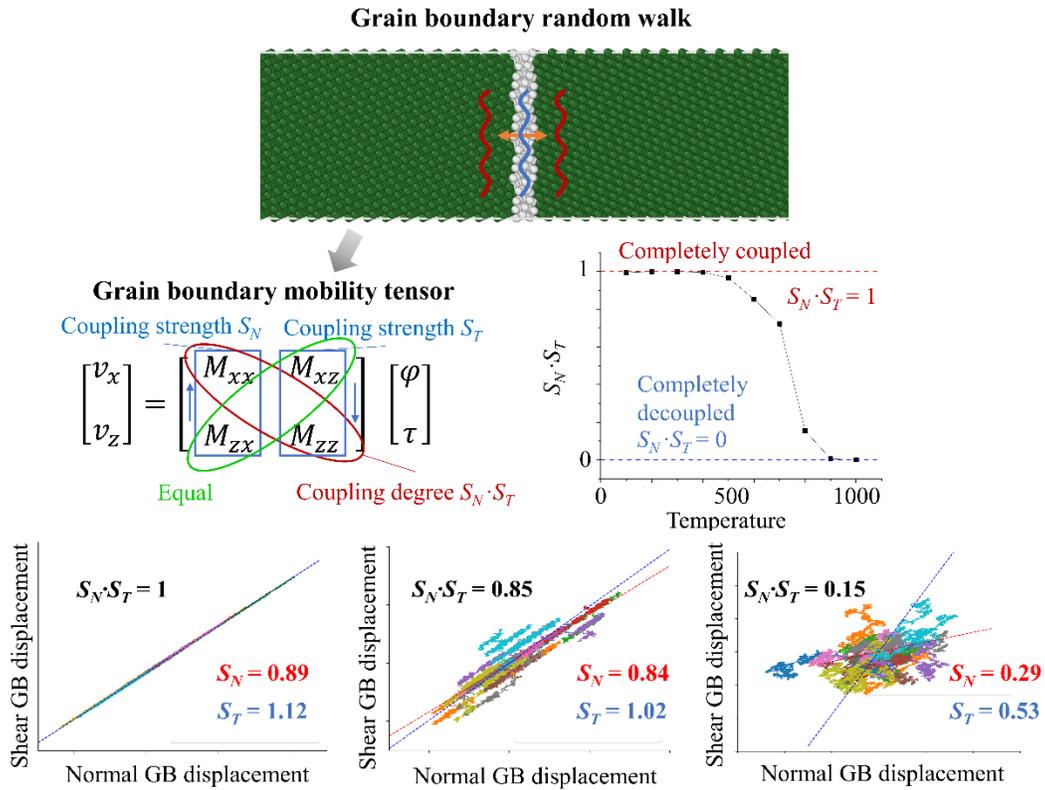

**Abstract**

Grain boundary (GB) mobility has been conventionally computed as a single value; however, a recent study has suggested that GB mobility should be expressed as a tensor. In this work, by using atomistic simulations, the concept of GB mobility being applied to the shear direction was re-examined and it is found that it follows the same physical rule as the conventionally defined GB mobility based on the normal direction. The interface random walk method was then used to compute the intrinsic GB mobility tensor at the zero-driving force limit. In order to compute the off-diagonal elements of the intrinsic GB mobility tensor, a shear coupling strength *S* is introduced in this study, which we believe can better reflect the intrinsic characteristics of a GB for its coupling trend between the normal and shear motion than the widely used shear coupling factor *β*. Furthermore, the effect of temperature and external driving force on the GB mobility tensor, especially on its symmetry, was systematically investigated. It is shown that the GB mobility in either the normal or shear direction can show a non-Arrhenius type dependence on the applied driving force, which is similar to the widely reported non-Arrhenius (or anti-thermal) dependence of GB mobility on temperature. Accordingly, the classical GB migration equation was adapted to describe the diverse variation of GB mobility due to changes in both temperature and driving force.

**1. Introduction**

Grain boundary (GB) mobility is one of the most important properties that influence the microstructural changes in polycrystalline materials during heat treatment or mechanical deformation. In the past, GB mobility *M* has been linked to grain growth thus considering only the normal migration of the GB plane, which can be calculated by the following equation [1]

$$M = v_\perp/P \qquad (1)$$

where $v_\perp$ is the GB normal migration velocity, and *p* is the external driving force. However, many GBs show the shear coupling effect [2], which means GBs can have a shear displacement tangent to the GB plane during its normal migration. Due to the shear coupling effect, normal GB migration can also be induced by shear stress, suggesting the commonality between the shear and normally induced GB migration. Recently, the shear and normally induced GB migration and their underlying mechanisms have been unified under the framework based on disconnection nucleation and migration [1]. It is thus natural to extend the concept and definition of GB mobility based on Eq. (1) to the shear direction of a GB. However, GB shear mobility has been barely discussed in previous studies until Chen et al. [3] proposed the concept of GB mobility tensor in 2020:

$$\begin{bmatrix} v_x \\ v_y \\ v_z \end{bmatrix} = \begin{bmatrix} M_{xx} & M_{xy} & M_{xz} \\ M_{yx} & M_{yy} & M_{yz} \\ M_{zx} & M_{zy} & M_{zz} \end{bmatrix} \begin{bmatrix} \varphi \\ \tau_y \\ \tau_z \end{bmatrix} \qquad (2)$$

Here $x$ direction is perpendicular to the GB plane, $y$ and $z$ directions are parallel to the GB plane, $\varphi$ is the normal driving force such as that caused by an energy jump across the GB [4,5], and $\tau$ is the shear stress tangent to the GB plane. By extending the concept of GB mobility to the shear direction, it is expected that the GB mobility tensor can be used to characterize not only GB growth but also other types of GB-mediated microstructural evolution, such as GB sliding [6] and grain rotation [7–9] which are widely reported to result in softening and dominate the plasticity in polycrystalline materials when the grain size falls below the reverse-Hall Petch limit.

Chen et al. [3] computed the GB mobility tensor by Eq. (2) and showed that the GB mobility tensor should be symmetric according to the Onsager relation. However, in their study, some of the used driving force (e.g. shear stress $\tau$ of 200 MPa) is larger than typical experimental situations (typically $10^2 - 10^6$ Pa [10]). This may have changed the GB migration mechanism [11] and accordingly, may have not reflected the intrinsic property of the GB mobility tensor. Therefore, the intrinsic GB mobility, which refers to the ability of a GB to migrate under infinitesimal driving force, needs to be accurately determined for all components of the GB mobility tensor so that its symmetry can be strictly validated.

Trautt et al. [12] have proposed the interface random walk method to extract the intrinsic normal GB mobility at the zero-driving force limit from the Einstein relation:

$$\langle h(t)^2 \rangle = \frac{2M_{xx} k_B T}{A} t \qquad (3)$$

where $k_B$ is Boltzmann constant, $T$ is temperature, $A$ is the GB area, and $<h(t)^2>$ is the mean square displacement (MSD) of the average normal migration of the whole GB plane at finite temperatures. Later, Karma et al. [13] extended this method to calculate the grain sliding mobility along the horizontal direction (e.g., $z$ direction):

$$\langle X(t)^2 \rangle = \frac{2M_{zz}k_BT}{A}t \tag{4}$$

where $<X(t)^2>$ is the MSD of the tangential displacement of grains at finite temperatures. The results for sliding along the $y$ direction can be computed in a similar way. For demonstration purposes, the current study will be mainly based on the shear direction along $z$ and the coupling between $x$ and $z$ directions. However, it is yet to be validated if the grain sliding mobility as proposed by Karma et al. [13] in Eq. (4) is the same as the GB mobility along the shear directions as defined by Chen et al. [3] in Eq. (2). Even if it is true, the interface random walk method can only be used to determine the diagonal components in the GB mobility tensor such as the $M_{xx}$ and $M_{zz}$. Therefore, new methods need to be developed for the computation of the off-diagonal components such as $M_{xz}$ and $M_{zx}$ at the zero-driving force limit.

Moreover, in the past, the shear coupling factor is the most commonly used parameter to describe the correlation between the GB motion along the normal and shear directions which is defined as $\beta = v_z/v_x$ based on the GB migration velocity, or by $\beta = M_{zx}/M_{xx}$ (or $M_{zz}/M_{xz}$) based on the GB mobility tensor as proposed by Chen et al. [3]. The previous study shows that the shear coupling factor in a GB may differ significantly if the GB motion is driven differently, e.g., by energy jump $\varphi$ or by a shear stress $\tau$ [14] even if it is measured at the same temperature. Such observation has led to the conclusion that GB shear coupling is not a GB property [14]. For example, at temperatures close to the melting point, one may expect the shear coupling factor to either approach 0 ($\beta \to 0$) or infinity ($\beta \to \infty$), depending on if the GB motion is driven in the normal or shear directions. The contrasting values make it difficult to describe the shear coupling ability of a GB based on the conventional shear coupling factor, and it is somewhat misleading that a larger shear coupling factor does not always mean a stronger shear coupling trend in the GB. The definition of the conventional shear coupling factor, therefore, needs to be modified, or alternatively, a new parameter is needed, so that it can consistently and unambiguously describe the ability of a GB to shear couple under arbitrary types of driving force.

In addition, the conventional normal GB mobility has been found to show many interesting phenomena, one of the most noteworthy is the so-called anti-thermal or non-Arrhenius behavior [15–17] as well as its dependency on temperature and driving force [11,18]. Will the GB mobility along the shear direction follow the same physical laws, or fundamentally, can the concept of mobility be really applied to describe the GB motion along the shear direction? If so, how will temperature and driving force influence the GB mobility tensor and its symmetry?

In this study, we tend to address these fundamental questions regarding the GB motion along the shear directions and find a way to compute the intrinsic GB mobility tensor at the zero-driving force limit. Based on that, we systematically study its symmetry and dependence on external factors such as temperature and driving forces.

## 2. Methods

The twist Σ15 (2 1 1) Ni GB (P14 in the Olmsted database[19], as shown in Fig. 1) is used as the main model system in the current study. The simulation cell is periodic in the *y* and *z* directions and non-periodic (shrink-wrapped) in the *x* direction. There are some considerations for using the Ni Σ15 (2 1 1) GB: first, based on a previous study [11], this GB shows strong anti-thermal behavior and has obvious GB migration phenomenon at temperatures as low as 100 K under a relatively small driving force. It is therefore a good system to study the effects of temperature and external driving force on the GB mobility tensor. Second, the Ni Σ15 (2 1 1) GB shows almost zero mobility in the *y* direction, and the shear coupling effect is found only in the *z* direction. Accordingly, the 3×3 GB mobility tensor (Eq. (2)) can be simplified into a 2×2 tensor:

$$\begin{bmatrix} v_x \\ v_z \end{bmatrix} = \begin{bmatrix} M_{xx} & M_{xz} \\ M_{zx} & M_{zz} \end{bmatrix} \begin{bmatrix} \varphi \\ \tau \end{bmatrix} \quad (5)$$

It is expected that the methods proposed in this study based on the 2×2 tensor can be readily extended to study the more general 3×3 GB mobility tensor when shear coupling exists in more than one shear direction.

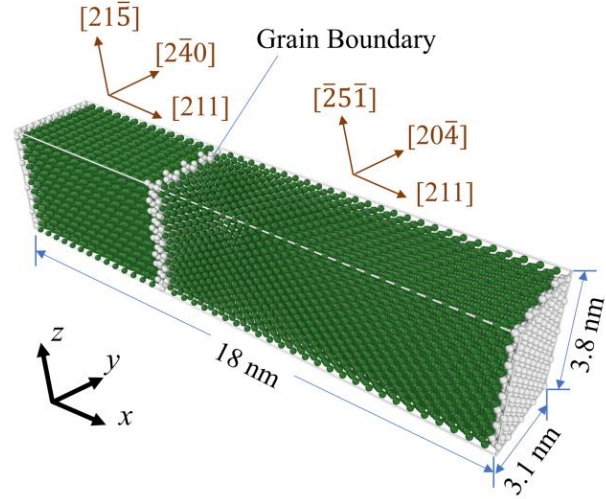

Figure 1 The atomistic model of Ni Σ15 (2 1 1) GB

The simulations were carried out through the Large-scale Atomic/Molecular Massively Parallel Simulator (LAMMPS) package[20] with a timestep of 1 fs. The model was first expanded at different temperatures with the corresponding thermal expansion coefficient and then equilibrated under isothermal-isobaric ensemble (NPT) for 10 ps, followed by a short (5 ps) annealing under microcanonical ensemble (NVE) with Berendsen thermostat[21].

For the random walk simulations, the equilibrated model at finite temperatures (100–1000K) fluctuated randomly under the NVE ensemble for 5 ns. The order parameter as defined in Ref. [5] was used to track the normal GB displacement in the x direction, $<h(t)^2>$. The GB displacements in the y and z direction, $<X(t)^2>$, were calculated by the relative shear displacement of the center of mass of a thin slab (~ 1 nm thick) at the top and bottom of the model, respectively, in x direction. The simulations at each temperature were repeated 20 times with different random seeds for initial velocity distribution. During the post-data processing, each 5 ns simulation was split into $10 \times 500$ ps simulations. In total, 200 sets of simulation data for each temperature were obtained. As an example, Fig. 2 shows that even at a low temperature of 200 K, the position of the Σ15 (2 1 1) Ni GB fluctuated intensely and randomly (the average position remained at 0), and there is significant displacement only in the x and z directions (Fig. 2(a) and (c)) but not in the y direction, suggesting a close-to-zero mobility along the y direction ($M_{yy} \approx 0$).

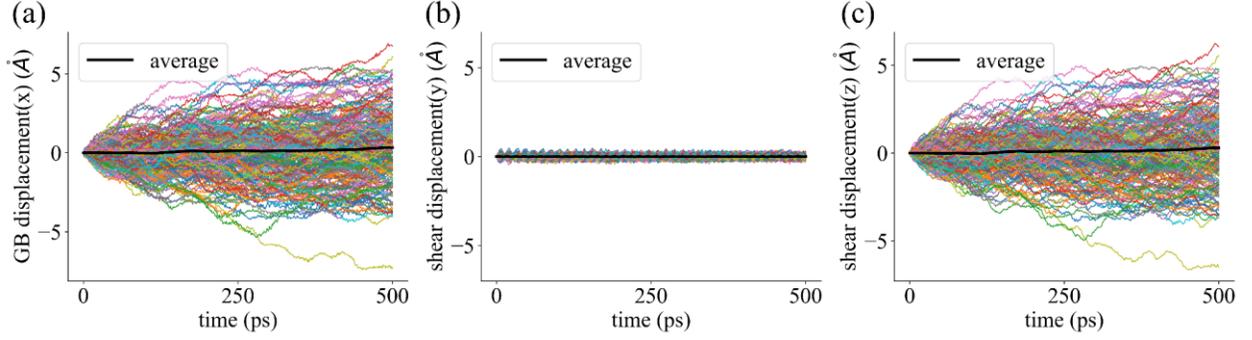

Figure 2 The fluctuation of the average position of Ni Σ15 (2 1 1) GB in (a) *x*, (b) *y*, and (c) *z* directions at 200 K.

The cumulative distribution function *f(x)* of the GB displacement for each time step *t* can be fitted as [22]

$$f(x) = \frac{1}{2}\left[1 + \text{erf}\left(\frac{x-\mu}{\sigma\sqrt{2}}\right)\right] \qquad (6)$$

where *f(x)* is the possibility that *h(t)* or *X(t)* falls in the range (∞, *x*], erf is the error function and the fitted $\sigma^2$ is the mean square displacement of the GB $\langle h(t)^2 \rangle$ or $\langle X(t)^2 \rangle$. Then, the diagonal components in the GB mobility tensor, $M_{xx}$ and $M_{zz}$, are calculated by using Eqs. (3) and (4), respectively.

To extract the shear coupling information from the random walk simulation, we linearly fitted the *X(t)* vs. *h(t)* and *h(t)* vs. *X(t)* data using the least square error method

$$Error = \sum_{i=0} |P(x_i) - y_i|^2 \qquad (7)$$

where $y_i$ is the original data, and *P(x)* is the fitted line. We define the slope of the line fitting the *X(t)* vs. *h(t)* and *h(t)* vs. *X(t)* results as the normal and tangential shear coupling strength $S_N$ and $S_T$, respectively. Based on the definition, the shear coupling strength reflects the ability for a GB to move in one direction due to shear coupling when it is driven along the other direction, i.e., $S_N$ (or $S_T$) represents the ratio of the coupled GB displacement along the shear (or normal) direction to that caused by a normal (or shear) driving force. At the first glance, one may suspect that the shear coupling strength seems no different from the conventional shear coupling factor and $S_N$ and $S_T$ seem to be related by a simple reciprocal relationship. It will be shown in the following sections that there are some circumstances that the shear coupling strength $S$ has clear advantages over the shear coupling factor $\beta$ and $S_N$ and $S_T$ are not always reciprocal to each other.

Simulations of GB migration driven by finite external driving forces were also carried out. The energy-conserving orientational (ECO) synthetic driving force method [5,23] was used as the normal driving force $\varphi$. For the shear stress $\tau$, two opposite shear forces were added on the two thin slabs at the ends of the model. The simulations were performed under the NVE ensemble with the Berendsen thermostat until the GB reached one end of the model (5 ns maximum). Each simulation was repeated 20 times with different random seeds for initial velocity distribution. The velocity of the GB migration was computed by fitting the slope of the cumulated GB displacement-time curve. The GB mobility tensor at finite driving forces can then be calculated from Eq. (5).

## 3. Results

*3.1 The intrinsic GB mobility tensor*

Fig. 3 shows that when the external driving force is small (e.g., less than 10 MPa in the current study), the $M_{xx}$ and $M_{zz}$ determined by the random walk method are equal to that determined by adding finite driving forces (for both $\varphi$ and $\tau$). These results directly address some of the fundamental questions raised in the Introduction section: The grain sliding mobility which was proposed by Karma et al. [13] is indeed the GB mobility in the shear direction as defined in the GB mobility tensor by Chen et al. [3] in Eqs. (2) and (5), and the GB motions in both the normal and shear directions at least follow the same physical law of Einstein relation. Besides, the shear and normal GB mobility $M_{xx}$ and $M_{zz}$ both show the anti-thermal behavior from 200 to 500 K, i.e., both $M_{xx}$ and $M_{zz}$ decrease with the increase in temperature. Other physical laws behind the anti-thermal behavior of the GB normal mobility proposed in the former studies [15,17,18,24] may also be applicable to the GB shear mobility, which will be discussed in a later section. So far, all analysis suggests that the concept of mobility can be applied to describe the GB motion along the shear direction.

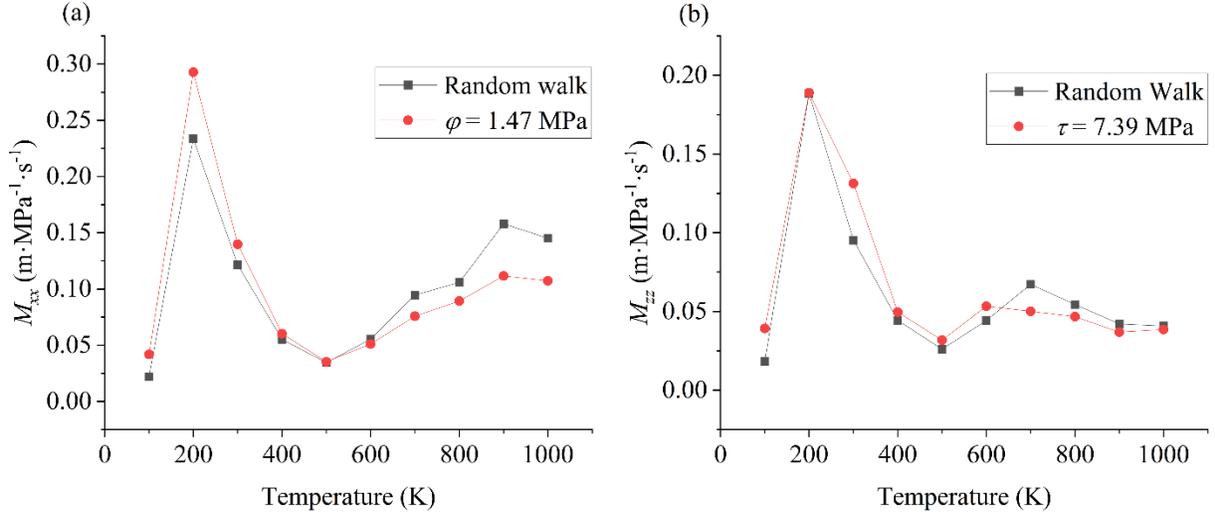

Figure 3 Comparison of $M_{xx}$ and $M_{zz}$ determined by interface random walk method to that by adding external driving forces (a) $\varphi$ and (b) $\tau$.

To extract the shear coupling information from random walk simulations, the shear coupling strength $S_N$ and $S_T$ as proposed in the Method section are computed respectively. Illustration of the determination of the shear coupling strength is shown in Fig. 4(g-i), in which the representative $X(t)$ vs. $h(t)$ curves obtained from the random walk simulations at 200, 600, and 800 K were shown. When the $X(t)$ and $h(t)$ exhibit a perfect linear relationship (strongly shear coupled, e.g., Fig. 4(g)), $S_N \cdot S_T =1$. If the linear relationship weakens, as shown in Fig. 4(h-i), both $S_N$ and $S_T$ decrease so that $S_N \cdot S_T$ becomes less than 1. Therefore, the shear coupling strength $S_N$ and $S_T$ as well as their product $S_N \cdot S_T$ are indicators of the linear correlation between the shear and normal GB migration, which reflects the degree of shear coupling in the GB. A more detailed discussion about the shear coupling strength and its product will be provided in Sections 3.2 and 4.1. Fig. 4(a-b) shows that $S_N$ is basically equal to the shear coupling factor $\beta_{zx}$ determined by $M_{zx}/M_{xx}$ from simulations under a small $\varphi$ (1.47 MPa), and similarly, $S_T$ is equal to the reciprocal of the shear coupling factor, $1/\beta_{xz}$, determined by $M_{xz}/M_{zz}$ from the simulations under a small $\tau$ (7.39 MPa). Therefore, the off-diagonal components in the GB mobility tensor can be computed directly from the random walk simulations, through

$$M_{zx} = M_{xx} \cdot S_N \qquad (8)$$

$$M_{xz} = M_{zz} \cdot S_T \qquad (9)$$

Although similar relations to Eq. (8) and (9) also exist in terms of the shear coupling factor, e.g., $M_{zx} = M_{xx} \cdot \beta_{zx}$ and $M_{xz} = M_{zz}/\beta_{xz}$, it should be noted that there is no easy way to compute two different shear coupling factors $\beta_{zx}$ and $\beta_{xz}$ at the zero-driving force limit. After all, there exists only one set of data, e.g., $<X(t)>$ vs. $<h(t)>$, from the interface random walk simulations, based on which only one shear coupling factor can be computed according to its definition, i.e., $\beta_{zx} = \beta_{xz} = <X(t)>/<h(t)>$, no matter what data processing technique may be used. This would be problematic if the shear coupling factors ($\beta_{zx}$ vs. $\beta_{xz}$) diverge from each other significantly at high temperatures. For example, as shown in Fig. 4(a) and (b), the $\beta_{zx}$ is close to zero while $\beta_{xz}$ approaches infinity at 1000 K, as expected.

Fig. 4(c) shows that the $M_{zx}$ and $M_{xz}$ determined by Eqs. (8) and (9) are equal, which proves the Osagar symmetry of the GB mobility tensor at the zero-driving force limit. In the following sections, we will discuss the effect of external factors, such as temperature and driving force, on the GB mobility tensor and its symmetry.

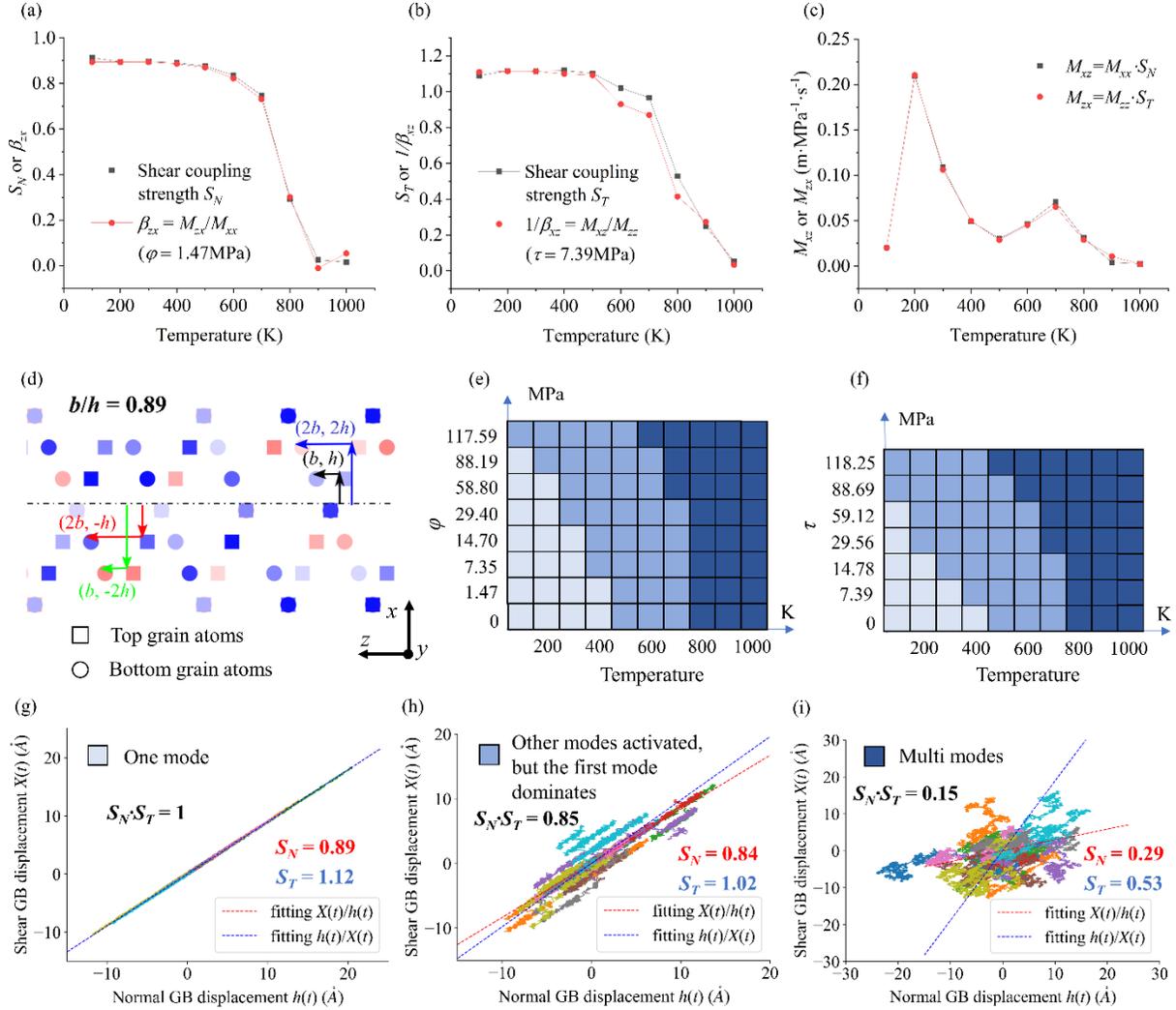

Figure 4 (a-b) Comparison between the shear coupling strength $S_N$ and $S_T$ determined from the interface random walk method and the shear coupling factor $\beta_{zx}$ and $1/\beta_{xz}$ determined by adding external driving forces. (c) Comparison between $M_{zx}$ and $M_{xz}$ computed by $M_{xx} \cdot S_N$ and $M_{zz} \cdot S_T$, respectively, from the interface random walk simulations. (d) The dichromatic pattern of the Ni Σ15 (2 1 1) GB showing the four disconnection modes with relatively low activation energy. The squares and circles represent the atoms belonging to the two different grains and the symbols are color-coded by their atomic position along the y direction. (e-f) The number of disconnection modes being activated under different (e) $\varphi$ and (f) $\tau$ (the data at 0 MPa is determined from interface random walk simulations) at different temperatures. (g-i) Illustration of the activated disconnection modes in (e-f) and the shear coupling strength from interface random walk simulations at (g) 200K, (h) 600K, and (i) 800K, respectively. Different colors represent data from the 20 independent interface random walk simulations.

*3.2 Effect of temperature on the GB mobility tensor*

When the GB movements along the *x* and *z* directions are strongly coupled (T < 500K), i.e., the *X(t)* and *h(t)* exhibit a perfect linear relationship as shown in Fig. 4(g), it can be derived according to the Einstein relation (Eqs. (3) and (4)) that $M_{zz}/M_{xx} = <X(t)^2>/<h(t)^2> = S_N^2=1/S_T^2$. Therefore, according to Eqs. (8) and (9), the GB mobility tensor can be simplified to

$$M = \begin{bmatrix} M_{xx} & S_N M_{xx} \\ S_N M_{xx} & S_N^2 M_{xx} \end{bmatrix} \text{ or } \begin{bmatrix} S_T^2 M_{zz} & S_T M_{zz} \\ S_T M_{zz} & M_{zz} \end{bmatrix} \quad (10)$$

in which all components in the GB mobility tensor are related. Once one component in the GB mobility tensor is computed, all the other components can be calculated accordingly.

According to the disconnection theory [1,2,25], the GB migration is mediated by the nucleation and migration of disconnections, which are line defects with both dislocation and height components, i.e., a Burgers vector *b* and a step height *h*. For each disconnection mode, the corresponding shear coupling factor *β* can be determined by *β* = *b*/*h*. The dichromatic pattern of the Ni Σ15 (2 1 1) GB (Fig. 4(d)) shows the four disconnection modes with relatively low activation energies (i.e., the combination of relatively small *b* and *h* according to Ref. [1]). The one with the smallest *b* and *h* pair, and accordingly, the lowest activation energy, produces a shear coupling factor of *b*/*h* = 0.89, suggesting that this is the mode being activated at T < 500K at the zero-driving force limit. As shown in Fig. 4(e-f), with the increase of temperature, disconnection modes of higher activation energies are also activated (i.e., disconnection modes with *b*/*h* ≠ 0.89 based on the *X(t)* *vs.* *h(t)* curves), which is an expected trend. The strong coupling condition and accordingly, the perfect linear correlation between *X(t)* and *h(t)*, is no longer valid. Fig. 5(a) shows the variation of $M_{zz}$ and $M_{xx} \cdot S_N^2$ computed from the random walk simulations. At T < 500K, the value of $M_{zz}$ is equal to $M_{xx} \cdot S_N^2$, and all components in the GB mobility tensor are coupled as described in Eq. (10). The deviation of $M_{xx} \cdot S_N^2$ from $M_{zz}$ at T ≥ 500K, however, means that the GB mobility tensor gradually becomes decoupled, and Eq. (10) no longer holds. Here the "couple" and "decouple" can be understood in this way: when a GB motion is perfectly "coupled", the shear and normal GB motion are not independent at all, which means any normal GB migration will cause a certain shear GB migration and vice versa; in contrast, when the GB motion is completely "decoupled", the GB can move in its normal or shear direction freely without causing any degree of GB migration in the other direction. Of course, there exists a regime where the GB motion is partially coupled. Fig. 4(e-f) and Fig. 5(a) suggest that the activation of other disconnection modes (e.g., at T ≥ 500K for the Ni Σ15 (2 1 1) GB) is the signal for decoupling.

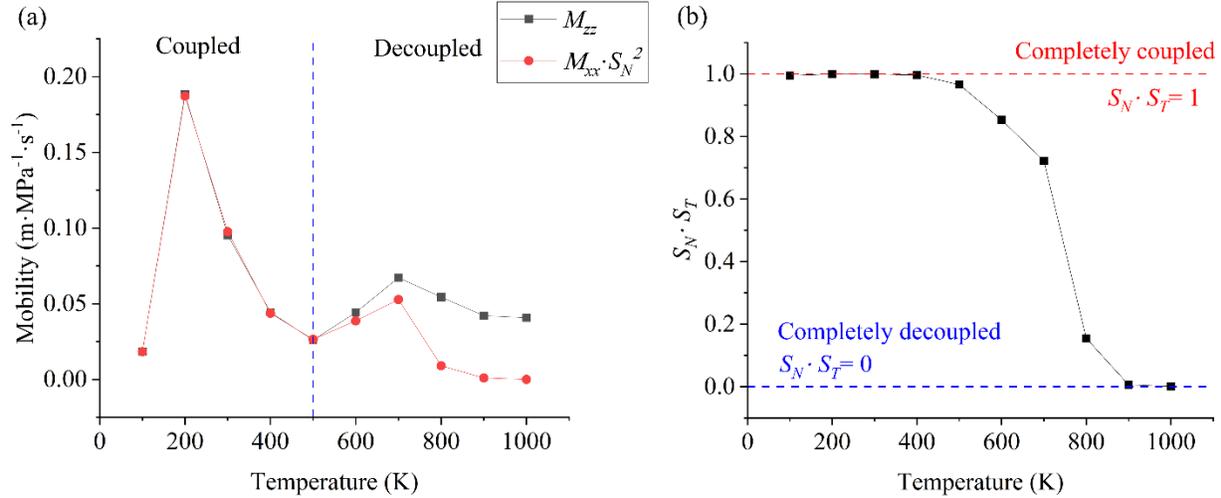

Figure 5 Variation of (a) $M_{zz}$ and $M_{xx} \cdot S_N^2$ and (b) $S_N \cdot S_T$ with temperature for Ni Σ15 (2 1 1) GB computed from interface random walk simulations.

To describe the transition from coupling to decoupling, a quantitative indicator is needed to measure the coupling degree of the GB mobility tensor. Fig. 5 shows that even though the GB mobility tensor is decoupled at T ≥ 500 K, the symmetry of the GB mobility tensor still holds at all temperatures (see Fig.4c). Therefore,

$$M_{xx} \cdot S_N = M_{zz} \cdot S_T \tag{11}$$

Multiplying both sides of Eq. (11) by $S_N$, we get

$$M_{xx} \cdot S_N^2 = S_N \cdot S_T \cdot M_{zz} \tag{12}$$

Comparing to Eq. (10) where $M_{xx} \cdot S_N^2 = M_{zz}$, the prefactor $S_N \cdot S_T$ in Eq. (12) can be used to describe the degree of coupling between the GB mobility along the normal and shear directions. When $S_N \cdot S_T = 1$, $M_{xx} \cdot S_N^2 = M_{zz}$, Eq. (10) holds, and the GB mobility tensor is completely coupled (T < 500K); when the GB mobility tensor becomes partially decoupled (T ≥ 500K), $S_N \cdot S_T$ starts to drop from 1; when $S_N \cdot S_T$ reaches 0 (T = 1000K), $M_{xx}$ and $M_{zz}$ become independent, and the GB mobility tensor is completely decoupled. The whole process is shown in Fig. 5(b).

*3.3 Effect of driving force on the GB mobility tensor*

To further study the effect of driving force on the GB mobility tensor, we computed the effect of both $\varphi$ and $\tau$ on the potential energy surface at 0 K by following the same procedure as used in [11]. Generally, the self-learning metabasin escape (SLME) approach [26–28] was first adopted to search the neighboring local minima around the current GB configuration. And then, nudged elastic band (NEB) method [29–31] was applied to calculate the energy barriers between the two configurations with adding the external driving forces($\varphi$ and $\tau$).

As shown in Fig. 6, the Σ15 (211) Ni GB performs two-step migration with two distinct energy barriers, which has been previously reported [11,32]. Both $\varphi$ and $\tau$ could lower the GB migration energy barriers in a similar way, which can facilitate the activation of disconnection modes of higher energy barriers as shown in Fig. 4(e, f), therefore, decoupling the GB mobility tensor at lower temperatures.

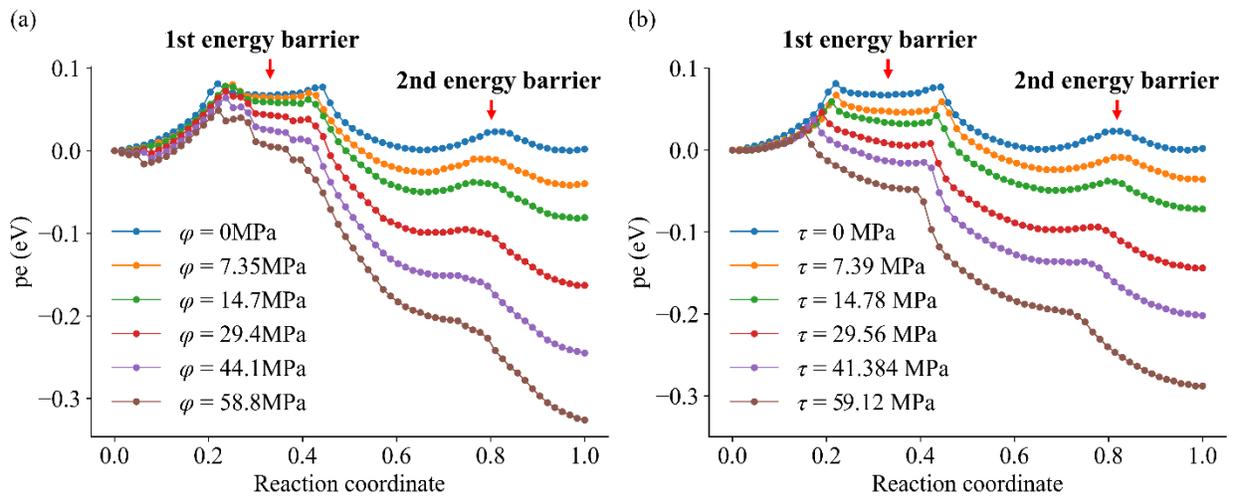

Figure 6 The effect of (a) $\varphi$ and (b) $\tau$ on the GB migration energy barrier at 0 K.

Furthermore, Fig. 7(a) shows that, when the external driving force exceeds 30Mpa, the symmetry of the GB mobility tensor is broken (Fig. 7(b)). This is expected since it has been widely reported previously that driving forces could strongly influence the value of GB mobility if it is computed by assuming a simple linear relation between GB velocity and driving force [1,10,11]. Large driving forces may cause a change in the underlying migration mechanism, e.g., from the diffusive to the ballistic regime [10]. Fig. 6 shows that when the driving force exceeds 30Mpa regardless of its type, the second peak of the energy barrier disappears, which explains why 30 Mpa is a critical value for the Ni Σ15 (2 1 1) GB: Above this value, a dramatic change in the migration mechanism may have occurred which breaks the symmetry of the GB mobility tensor.

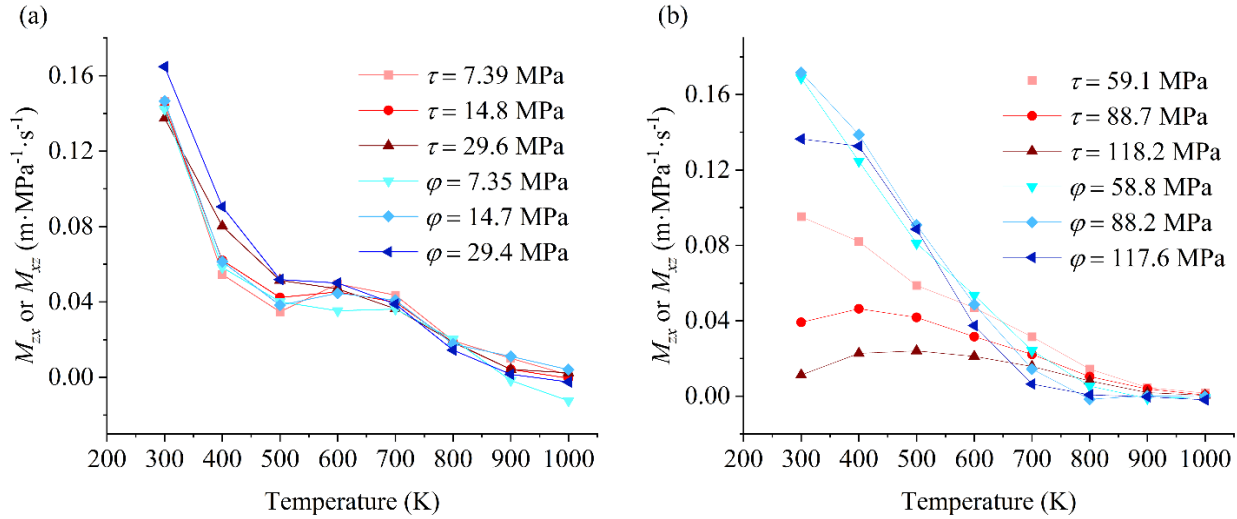

Figure 7 Comparison of $M_{zx}$ and $M_{xz}$ under different types and magnitudes of driving force.

In addition, there is also an "anti-thermal" trend of GB migration, or probably more accurately, an "anti-driving force" trend, with the increase of the external driving force. For example, as shown in Fig. 8, at a low temperature of 100 K, the GB mobilities in both directions, $M_{xx}$ and $M_{zz}$, first increase and then decrease with the increase of driving forces for both the $\varphi$ and $\tau$, which is similar to the anti-thermal phenomenon of GB mobility when the temperature changes under low driving force conditions (Fig. 3). In contrast, at a high temperature of 800 K, the GB mobilities are nearly constant and do not change with the driving force. A similar driving force dependent "anti-thermal" trend of GB mobility has also been reported previously, for example, by Deng and Schuh [10] and Race et. Al [33]. This phenomenon will be discussed in more details below. It needs to be clarified that the two data points in Fig. 8 at the two largest $\tau$ at 100 K as denoted by the red "★" symbols are computed based on the GB migration during which stagnation has occurred, which has significantly reduced the computed GB mobility value based on Eq. (2). The stagnation is different from the widely observed stick-slip behavior of GB migration [2], which may have been caused by subtle structural changes in the GB. This phenomenon is somewhat unexpected and will be the topic of a separate study. Nevertheless, the overall "anti-thermal" trend in GB mobility is not influenced even if the two data points with GB stagnation are excluded.

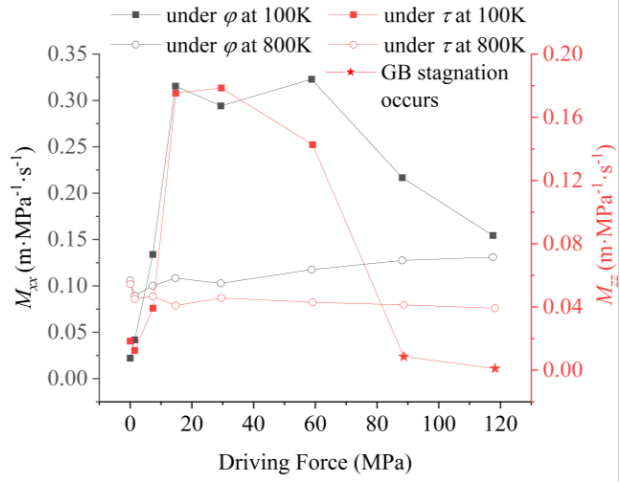

Figure 8 $M_{xx}$ and $M_{zz}$ vs. driving force at 100K and 800K, respectively.

## 4. Discussion

*4.1 The "Intrinsic" shear coupling factors and shear coupling strength*

The results in Fig. 4(a-b) can be used to justify why in the previous studies the GB shear coupling is not considered as a GB property if the goal is to use a single shear coupling factor as the indicator of the intrinsic shear coupling trend in a GB. For the specific case of the Ni Σ15 (2 1 1) GB, there actually exists a pair of two "intrinsic" shear coupling factors, instead of one, for each temperature at the zero-driving force limit, e.g., ($\beta_{zx}$, $\beta_{xz}$). Accordingly, the intrinsic GB mobility tensor can be expressed in terms of the intrinsic shear coupling factor pair as

$$\begin{bmatrix} v_x \\ v_z \end{bmatrix} = \begin{bmatrix} M_{xx} & M_{zz}/\beta_{xz} \\ \beta_{zx}M_{xx} & M_{zz} \end{bmatrix} \begin{bmatrix} \varphi \\ \tau \end{bmatrix} \quad (13)$$

where $\beta_{zx} = S_N$, and $\beta_{xz} = 1/S_T$. It demonstrates the necessity of representing GB mobility in the form of a tensor, which could better reflect the nature of shear coupling in a GB.

It needs to be emphasized that even though the shear coupling strength $S$ can be used to compute the shear coupling factor $\beta$ or vice versa at the zero-driving force limit, their physical meanings are different. The shear coupling factor $\beta$ is a parameter to measure the ratio of GB migration velocity or displacement in the shear direction to that in the normal direction, e.g., $\beta = v_z/v_x$. While the shear coupling strength $S$ can be used for the same purpose, it is also an indicator of the linear correlation strength between the GB displacements along the normal and shear directions. Accordingly, the shear coupling strength $S$ has the following advantages over the conventional shear coupling factor $\beta$.

First, the shear coupling strength $S$ is easier to compute than the shear coupling factor $\beta$. For example, when the GB randomly fluctuates at 800 K (Fig. 4 (i)), there are significant GB displacements in both shear and normal directions. If $\beta$ is used, one may find it difficult to determine its shear coupling status and compute an accurate $\beta$, since there is no easy way to differentiate the GB displacements caused by pure sliding (or pure migration) and by shear coupling. In contrast, the same fitting method as defined in Eq. (7) can be used to compute $S$ without the need of worrying about the migration status. The value of $S$ can then be used to determine the migration status as discussed below.

Second, the shear coupling strength $S$ can better reflect the change in the coupling status than the shear coupling factor $\beta$. For example, when the temperature approaches the melting point, the GB mobilities along different directions are decoupled. In this case, the shear coupling factor driven by $\tau$ approaches infinity ($\beta_{xz} \to \infty$), although it actually means there is no shear coupling at all. However, at temperatures close to the melting point, both $S_N$ and $S_T$ decrease and approach 0. The drop in the shear coupling strength therefore properly signals the drop in coupling trend during the GB migration.

Moreover, the product of the shear coupling strength $S_N \cdot S_T$ can describe the coupling degree of the GB mobility between the normal and shear directions (Eq. (12)). From $S_N \cdot S_T = 1$ to 0, the GB mobility tensor changes from completely coupled to completely decoupled for its component in the normal and shear directions (Fig. 5(b)). Accordingly, the change in $S_N \cdot S_T$ also reflects the individual contribution to the overall shear (or normal) displacement from that caused by pure sliding (or pure migration) and shear coupling, respectively. For instance, when $S_N \cdot S_T = 0$, the GB migration in the shear (normal) direction is caused by pure sliding (or pure migration).

Based on the above discussion, although the conventional shear coupling factor $\beta$ is fine to describe GB migration for most cases when finite driving forces are present, the shear coupling strength $S$ has clear advantages when the intrinsic shear coupling and mobilities of a GB needs to be evaluated. Accordingly, we believe the shear coupling can be considered as a GB property in terms of the shear coupling strength at the zero-driving force limit, which is embedded in the intrinsic GB mobility tensor.

*4.2 Adaptation of the classical GB migration equation*

According to the classical atom jump equation [17,34], the GB migration can be described as $N$ atoms jumping back and forth through the GB area. Therefore, the normal velocity of the GB migration $v$ can be given by[17]

$$\begin{aligned} v &= Nbv(\Gamma^+ - \Gamma^-) \\ &= Nbv\left[exp\left(\frac{-Q_1}{k_BT}\right) - exp\left(\frac{-(Q_1+Q_2)}{k_BT}\right)\right] \\ &= Nbvexp\left(\frac{-Q_1}{k_BT}\right)\left[1 - exp\left(\frac{-Q_2}{k_BT}\right)\right] \end{aligned} \qquad (14)$$

where $b$ is the atom jumping distance, $v$ is the attempt frequency, $\Gamma^+$ and $\Gamma^-$ is the probability of the atom jumping forth (+) and back (-), and $k_B$ is Boltzmann constant. $Q_1$ is the energy barrier of the GB migration and $Q_2$ is the energy slump introduced by the external driving force (as shown in Fig. 9). Therefore, the GB normal mobility $M_\perp$ can be expressed as

$$M_\perp = Nbvexp\left(\frac{-Q_1}{k_BT}\right)\left[1 - exp\left(\frac{-Q_2}{k_BT}\right)\right]/p \qquad (15)$$

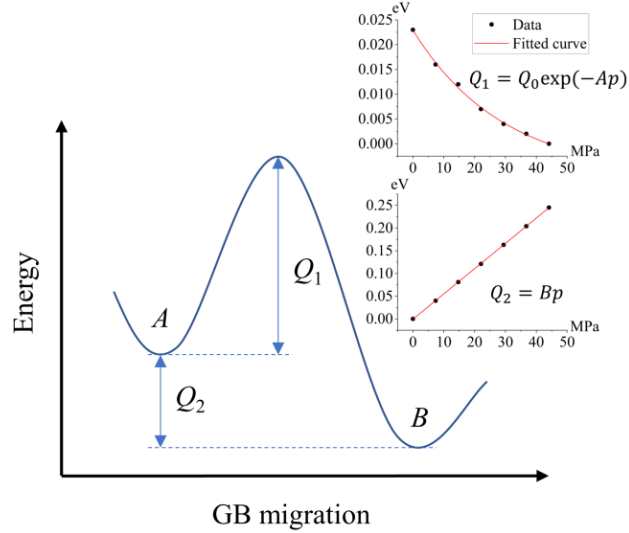

Figure 9 Illustration of the effect of driving force on the potential energy surface for GB migration.

Since there are two energy barriers in the Ni Σ15 (2 1 1) GB and the first energy barrier barely changes when an external driving force is applied, as shown in Fig. 6(a), the change in GB mobility should be mainly caused by the change in the second energy barrier. Therefore, we fit the $Q_1$ vs. $\varphi$ by using the second energy barrier. The insets of Fig. 9 show that $Q_1$ and $Q_2$ vs. $\varphi$ can be fitted using the following relations

$$Q_1 = Q_0 exp(-Ap) \tag{16}$$

$$Q_2 = Bp \tag{17}$$

where $Q_0$ is the initial GB migration energy barrier, and $A$ and $B$ are fitting constants. When the driving force is small enough, $Q_1 \rightarrow Q_0$ and because $Q_2 \ll k_BT$, $[1 - exp(-Q_2/k_BT)] \rightarrow Q_2$ (first-order Maclaurin series expansions). Therefore, we can reduce Eq. (15) to

$$M_\perp = \frac{C}{k_BT} exp\left(\frac{-Q_0}{k_BT}\right) \tag{18}$$

where $C$ is $Nbv$ which is a constant. Eq. (18) could explain the anti-thermal behavior of GB normal mobility, as shown in Fig. 3(a) and 11(a), which have been extensively discussed in previous studies[11,17]. It is worth noting that the energy barriers shown in Fig. 6 are based on NEB analysis at $T = 0$ where only one disconnection mode is activated, and accordingly, Eq. (18) (Fig. 10(a)) is based on the premise that there is only one disconnection mode. For Σ15 (2 1 1) Ni GB, multi modes are activated at 500 K, so the constant $C$ in Eq. (18) as well as the energy barriers should have changed, which explains the turning point of GB mobility at 500 K in Fig. 3.

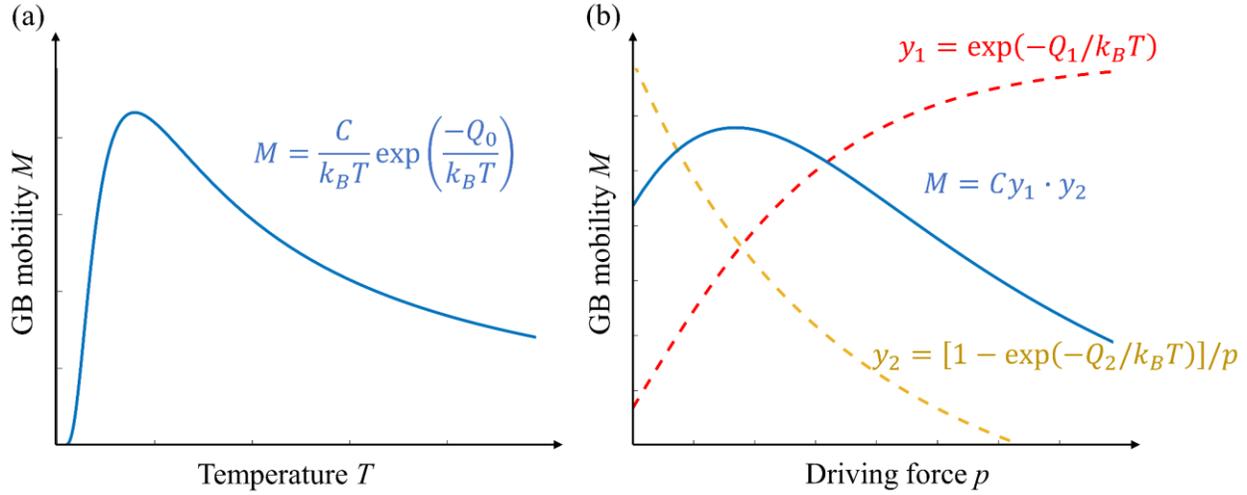

Figure 10 Variation of GB mobility as a function of (a) temperature $T$ and (b) driving force $p$.

When the driving force is large and temperature is low, i.e., $Q_2 \ll k_B T$ no longer holds, substituting Eqs. (16) and (17) into Eq. (15), we can see that the GB mobility first increases and then decreases (Fig. 10(b)), which is similar to the phenomenon shown in Fig. 8. When the temperature is high, $Q_1, Q_2 \ll k_B T$, therefore, GB normal mobility $M_\perp$ approaches a constant, which is also consistent with our finding in Fig. 8.

The fact that the GB shear mobility $M_\parallel$ shows the same "anti-thermal" trend with both temperature (Fig. 3(b)) and driving force (Fig. 8) as the GB normal mobility $M_\perp$ demonstrates that the GB shear migration also follows the same physical law. This is because the $\tau$ essentially causes the same effect on the energy barrier as the $\varphi$, as shown in Fig. 7, and as a result, the above analysis (Eqs. 14-18, Fig. 9-10) also applies to the shear direction. The effect of both temperature and driving force along both the shear and normal directions on the GB mobility tensor can be, therefore, unified and expressed by using Eq. (15).

## 5. Conclusions

In this study, we propose a method to compute the intrinsic GB mobility tensor at the zero-driving force limit through interface random walk simulation. Based on that, the influence of temperature and driving force on the GB mobility tensor is systematically studied. The following conclusions can be drawn based on this study:

- The concept of mobility being applied to the shear direction has been validated and it is found that the GB mobilities along both the normal and shear directions follow the same physical laws such as the Einstein relation and the classical GB migration equation.
- The shear coupling strength $S_N$ and $S_T$ proposed in the current study could be extracted from the interface random walk simulations and be used to compute the off-diagonal elements of the GB mobility tensor at the zero-driving force limit. Furthermore, the shear coupling strength can better reflect the shear coupling characteristics of the intrinsic GB migration behavior.
- At the low temperature-low driving force regime when only one disconnection mode is activated, the GB mobility tensor is completely coupled, i.e., each component in the GB mobility tensor is correlated, and once one component is determined, the others could be determined accordingly based on the shear coupling strength. With the increase in temperature or driving force, the GB mobility tensor would go through a coupled-to-decoupled transition during which the normal and shear GB migration gradually becomes independent of each other. The signal of decoupling is the activation of disconnection modes with higher energy barriers than its ground state and the parameter $S_N \cdot S_T$ (from 1 to 0) could reflect the coupling degree of the mobility tensor.
- The symmetry of the GB mobility tensor is strictly validated under small driving forces (< 30 MPa), but it fails under large driving forces (>30 MPa). This is consistent with previous findings that large driving forces could fundamentally change the underlying GB migration mechanisms.
- Both normal driving force $\varphi$ and shear stress $\tau$ could lower the GB migration energy barrier in a similar way and result in a similar "anti-thermal" trend of GB mobility with the driving force. The classical GB migration equation can be adapted to describe the "anti-thermal" trend of GB mobility due to changes in both temperature and driving force.

## Acknowledgment

The authors thank Dr. David L Olmsted for sharing the 388 Ni GB structure database, and Dr. Penghui Cao for sharing the SLME simulation code. This research was supported by NSERC Discovery Grant (RGPIN-2019-05834), Canada, and the use of computing resources provided by Compute/Calcul Canada.
## References

[1] J. Han, S.L. Thomas, D.J. Srolovitz, Grain-boundary kinetics: A unified approach, Prog. Mater. Sci. 98 (2018) 386–476.
[2] J.W. Cahn, Y. Mishin, A. Suzuki, Coupling grain boundary motion to shear deformation, Acta Mater. 54 (2006) 4953–4975.
[3] K. Chen, J. Han, X. Pan, D.J. Srolovitz, The grain boundary mobility tensor, Proc. Natl. Acad. Sci. 117 (2020) 4533–4538.
[4] K.G. Janssens, D. Olmsted, E.A. Holm, S.M. Foiles, S.J. Plimpton, P.M. Derlet, Computing the mobility of grain boundaries, Nat. Mater. 5 (2006) 124–127.
[5] F. Ulomek, C.J. O'Brien, S.M. Foiles, V. Mohles, Energy conserving orientational force for determining grain boundary mobility, Model. Simul. Mater. Sci. Eng. 23 (2015) 025007.
[6] L. Wang, Y. Zhang, Z. Zeng, H. Zhou, J. He, P. Liu, M. Chen, J. Han, D.J. Srolovitz, J. Teng, Tracking the sliding of grain boundaries at the atomic scale, Science. 375 (2022) 1261–1265.
[7] L. Wang, J. Teng, P. Liu, A. Hirata, E. Ma, Z. Zhang, M. Chen, X. Han, Grain rotation mediated by grain boundary dislocations in nanocrystalline platinum, Nat. Commun. 5 (2014) 1–7.
[8] Y.B. Wang, B.Q. Li, M.L. Sui, S.X. Mao, Deformation-induced grain rotation and growth in nanocrystalline Ni, Appl. Phys. Lett. 92 (2008) 011903.
[9] Z. Kou, Y. Yang, L. Yang, X. Luo, B. Huang, Observing the dynamic rotation and annihilation process of an isolated nanograin at the atomic scale in Al, Mater. Charact. 147 (2019) 311–314.
[10] C. Deng, C.A. Schuh, Diffusive-to-ballistic transition in grain boundary motion studied by atomistic simulations, Phys. Rev. B. 84 (2011) 214102.
[11] X. Song, C. Deng, Driving force induced transition in thermal behavior of grain boundary migration in Ni, ArXiv Prepr. ArXiv220808069. (2022).
[12] Z.T. Trautt, M. Upmanyu, A. Karma, Interface mobility from interface random walk, Science. 314 (2006) 632–635.
[13] A. Karma, Z.T. Trautt, Y. Mishin, Relationship between equilibrium fluctuations and shear-coupled motion of grain boundaries, Phys. Rev. Lett. 109 (2012) 095501.
[14] K. Chen, J. Han, S.L. Thomas, D.J. Srolovitz, Grain boundary shear coupling is not a grain boundary property, Acta Mater. 167 (2019) 241–247.
[15] J.L. Bair, E.R. Homer, Antithermal mobility in Σ7 and Σ9 grain boundaries caused by stick-slip stagnation of ordered atomic motions about Coincidence Site Lattice atoms, Acta Mater. 162 (2019) 10–18.
22